\documentclass[12pt,prd,tightenlines,nofootinbib,showpacs,showkeys]{revtex4}
\usepackage{bm}
\usepackage{graphics}
\usepackage{epsfig}
\begin{document}
\title{\begin{flushright}{\rm\normalsize SSU-HEP-09/8\\[5mm]}\end{flushright}
FINE AND HYPERFINE STRUCTURE\\ OF THE MUONIC $^3He$ ION}
\author{E.N. Elekina, A.P. Martynenko\footnote{E-mail:~mart@ssu.samara.ru}}
\affiliation{Samara State University, Pavlov Street 1, Samara 443011,
Russia}

\begin{abstract}
On the basis of quasipotential approach to the bound state problem
in QED we calculate the vacuum polarization, relativistic, recoil,
structure corrections of orders $\alpha^5$ and $\alpha^6$ to the
fine structure interval $\Delta E^{fs}=E(2P_{3/2})-E(2P_{1/2})$ and
to the hyperfine structure of the energy levels $2P_{1/2}$ and
$2P_{3/2}$ in muonic $^3_2He$ ion. The resulting values $\Delta
E^{fs}= 144803.15~~\mu eV$, $\Delta \tilde
E^{hfs}(2P_{1/2})=-58712.90~~\mu eV$, $\Delta \tilde
E^{hfs}(2P_{3/2})=-24290.69~~\mu eV$ provide reliable guidelines in
performing a comparison with the relevant experimental data.
\end{abstract}



\maketitle

\section{Introduction}

Simple atoms play important role in the check of quantum
electrodynamics (QED), the bound state theory and precise
determination of fundamental physical constants (the fine structure
constant, the lepton and proton masses, the Rydberg constant, the
proton charge radius, etc) \cite{EGS,SGK,MT}. Light muonic atoms
(muonic hydrogen $(\mu p)$, muonic deuterium, ions of muonic helium
etc.) are distinguished among simple atoms by the strong influence
of the vacuum polarization (VP) effects, recoil effects, nuclear
structure and polarizability effects on the structure of the energy
levels. The comparison of the theoretical value of the fine and
hyperfine splittings in muonic helium ions with the future
experimental data will lead to a more precise value of the helion
charge radius and the check of quantum electrodynamics with the
accuracy $10^{-7}$. The energy levels of muonic helium ions were
theoretically studied many years ago in \cite{b1,b2,b3,Romanov,GD}
both on the basis of the relativistic Dirac equation and
nonrelativistic approach, accounting different corrections by the
perturbation theory (PT). In these papers the basic contributions to
the energies for the $(2P-2S)$ transitions in muonic helium $(\mu
^3_2He)^+$ were evaluated with the accuracy $0.1~meV$.

In this work we continue the investigation \cite{m1} of the energy
spectrum of $(\mu ^3_2He)^+$ in the $P$-wave part. The aim of the
present study is to calculate such contributions of orders
$\alpha^5$ and $\alpha^6$ both in the fine and hyperfine structure
of the energy states $2P_{1/2}$, $2P_{3/2}$, which are connected
with the electron vacuum polarization, the recoil and structure
effects, the muon anomalous magnetic moment and the relativistic
corrections. The role of all these effects is crucial in order to
obtain high theoretical accuracy. Our purpose also consists in the
refinement of the earlier performed calculations in
\cite{b1,b2,b3,GD} and in the derivation of the reliable numerical
estimate for the structure of $P$-wave levels in the ion  $(\mu
^3_2He)^+$, which can be used for the comparison with experimental
data. Modern numerical values of fundamental physical constants are
taken from Ref.\cite{MT}: the electron mass
$m_e=0.510998910(13)\cdot 10^{-3}~GeV$, the muon mass
$m_\mu=0.1056583668(38)~GeV$, the fine structure constant
$\alpha^{-1}=137.035999679(94)$, the proton mass $m_p$ =
0.938272013(23)~GeV, the helion mass 2.808391383 (70)~GeV, the
helion magnetic moment $\mu_h=-2.127497723(25)$, the muon anomalous
magnetic moment $a_\mu=1.16592069(60)\cdot 10^{-3}$.

\section{Fine structure of $P$ - wave energy levels}

Our approach to the investigation of the energy spectrum of muonic
helium ion $(\mu ^3_2He)^+$ is based on the use of quasipotential
method in quantum electrodynamics \cite{m2,m3,m4}, where the
two-particle bound state is described by the Schr\"odinger equation.
The basic contribution to the muon and proton interaction operator
is determined by the Breit Hamiltonian \cite{t4,p1,p2}:
\begin{equation}
H=\frac{{\bf p}^2}{2\mu}-\frac{Z\alpha}{r}-\frac{{\bf p}^4}{8m_1^3}-
\frac{{\bf p}^4}{8m_2^3}+\frac{\pi Z\alpha}{2}\left(\frac{1}{m_1^2}+
\frac{1}{m_2^2}\right)-
\end{equation}
\begin{displaymath}
-\frac{Z\alpha}{2m_1m_2r}\left({\bf p}^2+\frac{{\bf r}({\bf rp}){\bf p}}
{r^2}\right)+\Delta V^{fs}(r)+\Delta V^{hfs}(r),
\end{displaymath}
where $m_1$, $m_2$ are the muon and proton masses, $\mu=m_1m_2/(m_1+m_2)$ is
the reduced mass, $\Delta V^{fs}$ is the muon spin-orbit interaction:
\begin{equation}
\Delta V^{fs}(r)=\frac{Z\alpha}{4m_1^2r^3}\left[1+\frac{2m_1}{m_2}+2a_\mu
\left(1+\frac{m_1}{m_2}\right)\right]({\bf L}{\mathstrut\bm\sigma}_1),
\end{equation}
$\Delta V^{hfs}$ is the helion spin-orbit interaction and the
interaction of the muon and helion spins. The leading order
$(Z\alpha)^4$ contribution to the fine structure is determined by
the operator $\Delta V^{fs}$. As it follows from Eq.(2), the
potential $\Delta V^{fs}$ includes also the recoil effects (the
Barker-Glover correction \cite{BG}) and the muon anomalous magnetic
moment $a_\mu$ correction. The fine structure interval
$(2P_{3/2}-2P_{1/2})$ for the ion $(\mu ^3_2He)^+$ can be written in
the form:
\begin{equation}
\Delta E^{fs}=E(2P_{3/2})-E(2P_{1/2})=\frac{\mu^3(Z\alpha)^4}{32 m_1^2}
\left[1+\frac{2m_1}{m_2}+2a_\mu\left(1+\frac{m_1}{m_2}\right)\right]+
\frac{5m_1(Z\alpha)^6}{256}-
\end{equation}
\begin{displaymath}
-\frac{m_1^2(Z\alpha)^6}{64m_2}+\frac{\alpha(Z\alpha)^6 \mu^3}{32\pi m_1^2}\left[\ln\frac{\mu(Z\alpha)^2}
{m_1}+\frac{1}{5}\right]+\alpha(Z\alpha)^4A_{VP}+\alpha^2(Z\alpha)^4B_{VP}.
\end{displaymath}

\begin{figure}[htbp]
\centering
\includegraphics{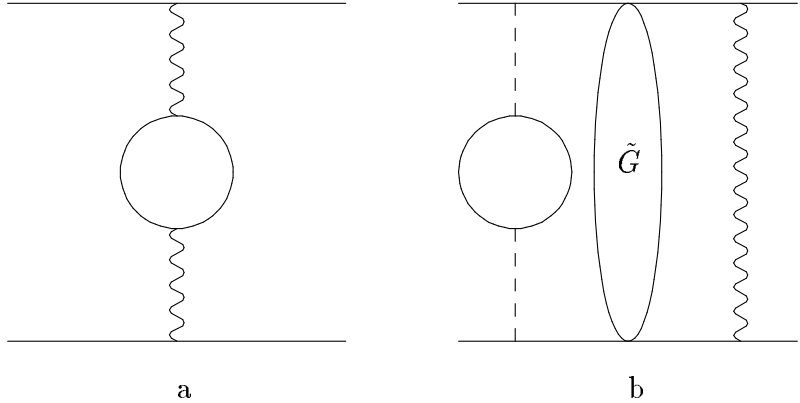}
\caption{One-loop vacuum polarization contributions to the fine and
hyperfine structure. The dashed line corresponds to the Coulomb
interaction. The wave line corresponds to the fine or hyperfine
interaction. $\tilde G$ is the reduced Coulomb Green's function.}
\end{figure}

This expression includes a relativistic correction of order
$(Z\alpha)^6$, which can be calculated with the aid of the Dirac
equation \cite{EGS,SY}, the correction of order $\alpha(Z\alpha)^6$
enhanced by the factor $\ln(Z\alpha)$ \cite{EY1,manakov,EY}, a
number of terms of fifth and sixth order in $\alpha$ which are
determined by the effects of the vacuum polarization. The
relativistic recoil effects of order $m_1(Z\alpha)^6/m_2$ in the
energy spectra of hydrogenic atoms were investigated in
Refs.\cite{EGS,VS1,VS2,SY,IBK}. In the fine splitting (3) they were
calculated in \cite{SY,IBK}. Additional corrections of the same
order were obtained in \cite{JP}. They do not depend on the muon
total momentum $j$ and give the contribution only to the Lamb shift.
The contributions to the coefficients $A_{VP}$ and $B_{VP}$ arise in
the first and second orders of perturbation theory. Numerical values
of the terms in the expression (3), which are presented in the
analytical form, are quoted in Table I for definiteness with the
accuracy $0.01~~\mu eV$. The fine structure interval (3) in the
energy spectrum of electronic hydrogen is considered for a long time
as a basic test of quantum electrodynamics \cite{SY,S1,R1972}.

The leading order vacuum polarization potential which gives the
contribution to the coefficient $A_{VP}$, is presented by the
Feynman diagrams in Fig.1. The one-loop vacuum polarization effects
lead to the modification both the Coulomb interaction and the
spin-orbit interaction in expressions (1), (2) \cite{t4,p1}:
\begin{equation}
\Delta V^C_{VP}(r)=\frac{\alpha}{3\pi}\int_1^\infty\rho(s)ds
\left(-\frac{Z\alpha}{r}\right)e^{-2m_esr},
\end{equation}
\begin{equation}
\Delta V^{fs}_{VP}(r)=\frac{\alpha}{3\pi}\int_1^\infty\rho(s)ds
\frac{Z\alpha}{4m_1^2r^3}\left[1+\frac{2m_1}{m_2}+2a_\mu\left(1+\frac{m_1}
{m_2}\right)\right]e^{-2m_esr}(1+2m_esr)({\bf
L}{\mathstrut\bm\sigma}_1),
\end{equation}
where the spectral function $\rho(s)=\sqrt{s^2-1}(2s^2+1)/s^4$,
$m_e$ is the electron mass. Averaging the potential (2) over the
wave functions of the $2P$ - state
\begin{equation}
\psi_{2P}({\bf
r})=\frac{1}{2\sqrt{6}}W^{5/2}re^{-\frac{Wr}{2}}Y_{1m}(\theta,\phi),~~~
W=\mu Z\alpha,
\end{equation}
we obtain the following contribution to the interval (3) (see
Fig.1(a)):
\begin{equation}
\Delta
E_1^{fs}=\frac{\mu^3(Z\alpha)^4}{32m_1^2}\left[1+\frac{2m_1}{m_2}+
2a_\mu\left(1+\frac{m_1}{m_2}\right)\right]\times
\end{equation}
\begin{displaymath}
\times\frac{\alpha}{3\pi}\int_1^\infty\rho(s)ds\int_0^\infty xdx
e^{-x\left(1+\frac{2m_es}{W}\right)}\left(1+\frac{2m_es}{W}x\right)=129.25~~\mu
eV.
\end{displaymath}
Although the integral in Eq.(7) can be calculated analytically, we
present here for simplicity only its numerical value.

Higher order corrections $\alpha^2(Z\alpha)^4$ entering in the
$a_\mu$ are taken into account in this expression as well as the
recoil effects. The same order contribution $\alpha(Z\alpha)^4$ can
be obtained in the second order perturbation theory (see Fig.1(b)).
In this case the energy spectrum is determined by the reduced
Coulomb Green's function \cite{Palchikov,hameka,p1}:
\begin{equation}
G_{2P}({\bf r},{\bf
r}')=-\frac{\mu^2(Z\alpha)}{36z^2z'^2}\left(\frac{3}{4\pi} {\bf
nn}'\right)e^{-\frac{z+z'}{2}}g(z,z'),
\end{equation}
\begin{equation}
g(z,z')=24z^3_{<}+36z^3_{<}z_{>}+36z^3_{<}z^2_{>}+24z^3_{>}+36z_{<}z^3_{>}+
36z^2_{<}z^3_{>}+49z^3_{<}z^3_{>}-
\end{equation}
\begin{displaymath}
-3z^4_{<}z^3_{>}-12e^z_{<}(2+z_{<}+z^2_{<})z^3_{>}-3z^3_{<}z^4_{>}+
12z^3_{<}z^3_{>}\left[-2C+Ei(z_{<}-\ln(z_{<})-\ln(z_{>})\right],
\end{displaymath}
where $z_{<}=min(z,z')$, $z_{>}=max(z,z')$, $C=0.577216...$ is the
Euler constant, $z=Wr$. Using Eqs. (8) and (9) we transform the
correction of order $\alpha(Z\alpha)^4$ to the fine structure in the
second order perturbation theory as follows:
\begin{equation}
\Delta E^{fs}_2=-\frac{\alpha(Z\alpha)^4\mu^3}{3456\pi m_1m_2}\left[1+2a_\mu+
(1+a_\mu)\frac{2m_1}{m_2}\right]\times
\end{equation}
\begin{displaymath}
\times\int_1^\infty\rho(s)ds\int_0^\infty dx
e^{-x\left(1+\frac{2m_es}{W}\right)}
\int_0^\infty\frac{dx'}{x'^2}e^{-x'}g(x,x')=140.56~~\mu eV.
\end{displaymath}

Let us consider the two-loop vacuum polarization contributions in
the one-photon interaction shown in Fig.2. They give the corrections
to the fine splitting of $P$- levels of order $\alpha^2(Z\alpha)^4$.

\begin{figure}[htbp]
\centering
\includegraphics{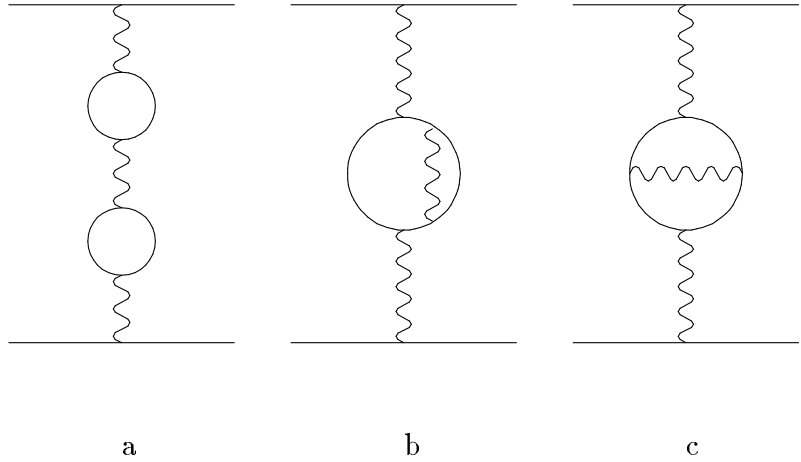}
\caption{Effects of two-loop electron vacuum polarization in the
one-photon interaction.}
\end{figure}

In order to obtain the particle-interaction operator for the
amplitude, corresponding to the diagram in Fig.2(a), it is necessary
to make the substitution
\begin{equation}
\frac{1}{k^2}\to\frac{\alpha}{3\pi}\int_1^\infty ds\frac{\sqrt{s^2-1}(2s^2+1)}
{s^4(k^2+4m_e^2s^2)}
\end{equation}
two times in the photon propagator. In the coordinate
representation, the interaction operator has the form \cite{m5,m6}:
\begin{equation}
\Delta V^{fs}_{VP-VP}(r)=\frac{Z\alpha}{r^3}\left[\frac{1+2a_\mu}{4m_1^2}+
\frac{1+a_\mu}{2m_1m_2}\right]({\bf L}{\mathstrut\bm\sigma}_1)\times
\end{equation}
\begin{displaymath}
\times\left(\frac{\alpha}{3\pi}\right)^2\int_1^\infty\rho(\xi)d\xi\int_1^\infty
\rho(\eta)d\eta\frac{1}{(\xi^2-\eta^2)}\left[\xi^2(1+2m_e\xi r)e^{-2m_e\xi r}-
\eta^2(1+2m_e\eta r)e^{-2m_e\eta r}\right].
\end{displaymath}
Averaging (12) over wave functions (6), we obtain the correction to
the interval (3):
\begin{equation}
\Delta E^{fs}_3=\frac{\mu^3\alpha^2(Z\alpha)^4}{72\pi^2}\left[\frac{1+2a_\mu}
{4m_1^2}+\frac{1+a_\mu}{2m_1m_2}\right]\int_1^\infty\rho(\xi)d\xi\int_1^\infty
\rho(\eta)d\eta\frac{1}{(\xi^2-\eta^2)}\times
\end{equation}
\begin{displaymath}
\times\int_0^\infty xdx
\left[\xi^2\left(1+\frac{2m_e\xi}{W}x\right)e^{-x\left(1+\frac{2m_e\xi}{W}\right)}
-\eta^2\left(1+\frac{2m_e\eta}{W}x\right)e^{-x\left(1+\frac{2m_e\eta}
{W}\right)}\right]= 0.20~~\mu eV.
\end{displaymath}
The two-loop vacuum polarization operator is needed to find the
interaction operator shown in Fig.2(b,c). The modification of the
photon propagator in this case has the form \cite{Eides}:
\begin{equation}
\frac{1}{k^2}\to \frac{2}{3}\left(\frac{\alpha}{\pi}\right)^2\int_0^1
\frac{f(v)dv}{4m_e^2+k^2(1-v^2)},
\end{equation}
\begin{equation}
f(v)=v\Biggl\{(3-v^2)(1+v^2)\left[Li_2\left(-\frac{1-v}{1+v}\right)+2Li_2
\left(\frac{1-v}{1+v}\right)+\frac{3}{2}\ln\frac{1+v}{1-v}\ln\frac{1+v}{2}-
\ln\frac{1+v}{1-v}\ln v\right]+
\end{equation}
\begin{displaymath}
\left[\frac{11}{16}(3-v^2)(1+v^2)+\frac{v^4}{4}\right]\ln\frac{1+v}{1-v}+
\left[\frac{3}{2}v(3-v^2)\ln\frac{1-v^2}{4}-2v(3-v^2)\ln v\right]+
\frac{3}{8}v(5-3v^2)\Biggr\}.
\end{displaymath}
The two-loop vacuum polarization potential and the correction to the
fine structure $(2P_{3/2}-2P_{1/2})$ are the following:
\begin{equation}
\Delta V_{2-loop,VP}^{fs}(r)=\frac{2\alpha^2(Z\alpha)}{3\pi^2r^3}
\left[\frac{1+2a_\mu}{4m_1^2}+\frac{1+a_\mu}{2m_1m_2}\right]\int_0^1
\frac{f(v)dv}{1-v^2}e^{-\frac{2m_er}{\sqrt{1-v^2}}}\left(1+\frac{2m_er}
{\sqrt{1-v^2}}\right)({\bf L}{\mathstrut\bm\sigma}_1),
\end{equation}
\begin{equation}
\Delta E^{fs}_4=\frac{\mu^3\alpha^2(Z\alpha)^4}{12\pi^2}
\left[\frac{1+2a_\mu}{4m_1^2}+\frac{1+a_\mu}{2m_1m_2}\right]\times
\end{equation}
\begin{displaymath}
\times\int_0^\infty x
dx\int_0^1\frac{f(v)dv}{1-v^2}e^{-x\left(1+\frac{2m_e}
{W\sqrt{1-v^2}}\right)}\left(1+\frac{2m_e}{W\sqrt{1-v^2}}x\right)=0.78~~\mu
eV.
\end{displaymath}

\begin{figure}[htbp]
\centering
\includegraphics{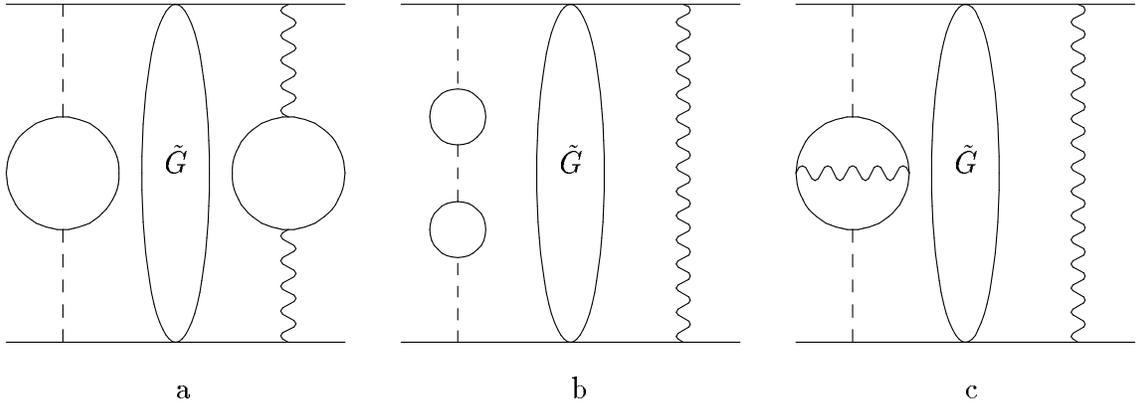}
\caption{Effects of two-loop electron vacuum polarization in the
second order perturbation theory. The dashed line corresponds to the
Coulomb interaction. The wave line corresponds to the fine or
hyperfine interaction. $\tilde G$ is the reduced Coulomb Green's
function.}
\end{figure}

Two-loop vacuum polarization contributions in the second order
perturbation theory shown in Fig.3, have the same order
$\alpha^2(Z\alpha)^4$. In order to calculate them, it is necessary
to employ relations (2), (4), (5), (8), and the modified Coulomb
potential by the two-loop vacuum polarization \cite{m2,m3}:
\begin{equation}
\Delta V^C_{VP-VP}(r)=\left(\frac{\alpha}{\pi}\right)^2\int_1^\infty
\rho(\xi)d\xi\int_1^\infty\rho(\eta)d\eta\left(-\frac{Z\alpha}{r}\right)
\frac{1}{\xi^2-\eta^2}\left(\xi^2e^{-2m_e\xi r}-\eta^2e^{-2m_e\eta r}\right),
\end{equation}
\begin{equation}
\Delta V^C_{2-loop,VP}=-\frac{2Z\alpha}{3r}\left(\frac{\alpha}{\pi}\right)^2
\int_0^1\frac{f(v)dv}{1-v^2}e^{-\frac{2m_er}{\sqrt{1-v^2}}}.
\end{equation}
The amplitude (a) in Fig.3 gives the following correction of order
$\alpha^2(Z\alpha)^4$ to the fine splitting:
\begin{equation}
\Delta E^{fs}_5=\frac{\mu^3\alpha^2(Z\alpha)^4}{1296\pi^2}\left[\frac{1+a_\mu}
{2m_1m_2}+\frac{1+2a_\mu}{4m_1^2}\right]\int_1^\infty\rho(\xi)d\xi
\int_1^\infty\rho(\eta)d\eta\times
\end{equation}
\begin{displaymath}
\times\int_0^\infty dx e^{-x\left(1+\frac{2m_e\xi}{W}\right)}
\int_0^\infty \frac{dx'}{x'^2}\left(1+\frac{2m_e\eta x'}{W}\right)
e^{-x'\left(1+\frac{2m_e\eta}{W}\right)}g(x,x')=0.28~~\mu eV.
\end{displaymath}
Two other contributions from the amplitudes (b), (c) in Fig.3 have
the similar integral structure. Their numerical values are included
in Table I. The summary result for the fine splitting $\Delta
E^{fs}$ in $(\mu ^3_2He)^+$ is presented here also. It takes into
consideration the numerous earlier performed calculations discussed
in the review article \cite{EGS} and new corrections obtained in
this work.

\begin{table}
\caption{Fine structure of $P$-wave energy levels in muonic $^3_2He$
ion.}
\bigskip
\begin{tabular}{|c|c|c|}   \hline
Contribution to the fine & Numerical value & Reference, \\
splitting $\Delta E^{fs}$            &of the contribution in
$\mu eV$ & equation  \\  \hline
Contribution of order $(Z\alpha)^4$ &              &         \\
$\frac{\mu^3(Z\alpha)^4}{32m_1^2}\left(1+\frac{2m_1}{m_2}\right)$ &
144186.48 & \cite{p1,b1}, (3) \\ \hline
Muon AMM contribution &       &           \\
$\frac{\mu^3(Z\alpha)^4}{16m_1^2}a_\mu\left(1+\frac{m_1}{m_2}\right)$
& 324.46 &\cite{p1,b1}, (3)  \\  \hline Contribution of order
$(Z\alpha)^6$: $\frac{5m_1(Z\alpha)^6}{256}$ &19.94     &
\cite{SY,IBK}, (3)  \\  \hline Contribution of order
$(Z\alpha)^6m_1/m_2$: $-\frac{m_1^2(Z\alpha)^6}{64m_2}$ &-0.60 &
\cite{SY,IBK}, (3)  \\  \hline
Contribution of order $\alpha(Z\alpha)^4$  &          &        \\
in the first order PT $\langle\Delta V^{fs}_{VP}\rangle$ & 129.25 &
\cite{p1,b1}, (7) \\   \hline
Contribution of one-loop muon  VP&          &        \\
in the first order PT $\langle\Delta V^{fs}_{MVP}\rangle$ & 0.01 &
\cite{p1,b1}, (7) \\   \hline
Contribution of order $\alpha(Z\alpha)^4$  &          &        \\
in the second order PT   & 140.56   & (10)   \\
$\langle\Delta V^C_{VP}\cdot\tilde G\cdot \Delta V^{fs}\rangle$ & &
\\   \hline
Contribution of order $\alpha(Z\alpha)^6$  &          & \cite{EGS,EY1,manakov}       \\
$\frac{\alpha(Z\alpha)^6\mu^3}{32\pi
m_1^2}\left[\ln\frac{\mu(Z\alpha)^2}{m_1} +\frac{1}{5}\right]$ &
-0.55 &  \\   \hline
VP Contribution in the second &      &      \\
order PT of order $\alpha^2(Z\alpha)^4$    & 0.28   & (20)   \\
$\langle\Delta V^C_{VP}\cdot\tilde G\cdot\Delta V^{fs}_{VP}\rangle$    &   &   \\  \hline
VP Contribution from $1\gamma$ interaction &      &      \\
of order $\alpha^2(Z\alpha)^4$ $\langle\Delta V^{fs}_{VP-VP}\rangle$
& 0.20 & (13)   \\   \hline
VP Contribution from $1\gamma$ interaction &      &      \\
of order $\alpha^2(Z\alpha)^4$ $\langle\Delta
V^{fs}_{2-loop,VP}\rangle$    & 0.78   & (17)   \\   \hline
VP Contribution in the second &      &      \\
order PT of order $\alpha^2(Z\alpha)^4$    & 0.03   & (18)   \\
$\langle\Delta V^C_{VP-VP}\cdot\tilde G\cdot\Delta V^{fs}\rangle$    &   &   \\  \hline
VP Contribution in the second &      &      \\
order PT of order $\alpha^2(Z\alpha)^4$    & 2.31   & (19)   \\
$\langle\Delta V^C_{2-loop,VP}\cdot\tilde G\cdot\Delta
V^{fs}\rangle$    & &   \\  \hline Summary contribution  & 144803.15
&
\\   \hline
\end{tabular}
\end{table}

\section{Hyperfine structure of the energy levels $2P_{1/2}$ and $2P_{3/2}$}

The leading order contribution to the hyperfine splitting of the
energy levels $2P_{1/2}$ and $2P_{3/2}$ in muonic helium ion $(\mu
^3_2He)^+$ of order $\alpha^4$ is determined by the potential (the
hyperfine part of the Breit potential) \cite{t4}:
\begin{equation}
\Delta
V_B^{hfs}(r)=\frac{\alpha\mu_h}{2m_1m_2r^3}\left[1+\frac{m_1}{m_2}-\frac{Zm_1m_p}
{2m_2^2\mu_h}\right]({\bf L}{\mathstrut\bm\sigma}_2)-
\end{equation}
\begin{displaymath}
-\frac{\alpha\mu_h(1+a_\mu)}{4m_1m_pr^3}\left[({\mathstrut\bm\sigma}_1
{\mathstrut\bm\sigma}_2)-3({\mathstrut\bm\sigma}_1{\bf n})
({\mathstrut\bm\sigma}_2{\bf n})\right],
\end{displaymath}
where ${\bf n}={\bf r}/r$. The operator (21) does not commute with
the operator of the muon total angular momentum ${\bf J}={\bf
L}+\frac{1}{2}{\mathstrut\bm\sigma}_1$. This leads to the mixing of
the $2P_{1/2}$ and $2P_{3/2}$ energy levels and, hence, to a more
complicated hyperfine structure of $P$-wave levels.

In order to calculate the diagonal matrix elements $\langle
2P_{1/2}|\Delta V_B^{hfs} |2P_{1/2}\rangle $ and $\langle
2P_{3/2}|\Delta V_B^{hfs}|2P_{3/2}\rangle $ we can use the following
replacements for the operators $({\bf s}_1{\bf s}_2)$ and $({\bf
L}{\bf s}_2)$, which involve the spin of the nucleus \cite{BS}:
\begin{equation}
{\bf s}_1\to{\bf J}\frac{\overline{({\bf s}_1{\bf J})}}{J^2},~~~
{\bf L}\to{\bf J}\frac{\overline{({\bf L}{\bf J})}}{J^2},
\end{equation}
where $\overline{({\bf s}_1{\bf J})}$, $\overline{({\bf L}{\bf J})}$
are eigenvalues of corresponding operators between the states with
equal orbital momentum $l$. In addition, the averaging over angles
in the second term in the right-hand side of (21) can be carried out
by means of the relation \cite{BS}:
\begin{equation}
\langle\delta_{ij}-3n_in_j\rangle=-\frac{1}{5}(4\delta_{ij}-3\overline{L_iL_j}-3\overline{L_jL_i}).
\end{equation}
The diagonal matrix elements $\langle 2P_{1/2}|\Delta
V^{hfs}|2P_{1/2}\rangle $ and $\langle 2P_{3/2}|\Delta
V^{hfs}|2P_{3/2}\rangle $ lead to the following hyperfine structure:
\begin{equation}
\Delta E^{hfs}(2P_{1/2})=E(2^3P_{1/2})-E(2^1P_{1/2})=
\end{equation}
\begin{displaymath}
=E_F\left[\frac{1}{3}+\frac{a_\mu}{6}+\frac{m_1}{6m_2}-\frac{Zm_1m_p}{12m_2^2\mu_h}+
\frac{m_1^3}{\mu^3}A_{rel}^{1/2}(Z\alpha)^2
+A_{VP}^{1/2}\alpha+B_{VP}^{1/2}\alpha^2\right],
\end{displaymath}
\begin{equation}
\Delta E^{hfs}(P_{3/2})=E(2^5P_{3/2})-E(2^3P_{3/2})=
\end{equation}
\begin{displaymath}
=E_F\left[\frac{2}{15}-\frac{a_\mu}{30}+\frac{m_1}{6m_2}-\frac{Zm_1m_p}{12m_2^2\mu_h}+
\frac{m_1^3}{\mu^3}A_{rel}^{3/2}(Z\alpha)^2
+A_{VP}^{3/2}\alpha+B_{VP}^{3/2}\alpha^2\right],
\end{displaymath}
where $E_F=Z^3\alpha^4\mu^3\mu_h/3m_1m_p$ is the Fermi energy for
the $n=2$ level. The calculation of the relativistic corrections
$A_{rel}^{1/2}$, $A_{rel}^{3/2}$ within this approach includes the
study of two-photon, three-photon exchange diagrams and the second
order perturbation theory contributions with the Breit Hamiltonian
determined by Eqs. (1), (2) and (21). More simple method for their
calculation is based on the relativistic Dirac equation
\cite{Breit,EY}. In that case, the hyperfine interaction potential
has the form:
\begin{equation}
\Delta V^{hfs}_D=e{\mathstrut\bm\mu}\frac{[{\bf r}\times
{\mathstrut\bm\alpha}]}{r^3}.
\end{equation}
Its contributions to the hyperfine splitting are given by
\begin{equation}
\Delta E^{hfs}_{rel}(2P_{1/2})=\frac{4\alpha\mu_h}{m_p}R_{1/2},
\end{equation}
\begin{displaymath}
\Delta E^{hfs}_{rel}(2P_{3/2})=-\frac{16\alpha\mu_h}{15m_p}R_{3/2},
\end{displaymath}
where the nuclear magnetic moment ${\mathstrut\bm\mu}=g_N\mu_N{\bf
s}_2$ ($\mu_N=e/2m_p$). The radial integrals $R_k=\int_0^\infty
g_kf_k dr$, which are characteristic  of this case, are determined
by the Dirac wave functions $f_k,g_k$ of the states $2P_{1/2}$ and
$2P_{3/2}$. Taking into account their explicit form \cite{BS}, we
obtain the following values of the relativistic corrections to the
hyperfine structure of the $P$ - wave levels:
\begin{equation}
A_{rel}^{1/2}=\frac{47}{72},~~~A_{rel}^{3/2}=\frac{7}{180}.
\end{equation}
These values of the coefficients coincide with the result obtained
in Ref.\cite{IBK} by analytically calculating the contribution of
order $m_1^2(Z\alpha)^6/m_2$ to the hyperfine structure of the $P$ -
wave levels of the hydrogen atom at $n=2$.

The fifth order contribution over $\alpha$ due to the electron
vacuum polarization (see diagrams (a), (b) in Fig.1) appears in the
hyperfine splitting in just the same way as in the fine structure of
the spectrum. The modification of the hyperfine part of the Breit
potential induced by the vacuum polarization is described by the
following expression (the substitution (11) is used) \cite{p1}:
\begin{equation}
\Delta
V_{VP}^{hfs}(r)=\frac{\alpha\mu_h}{2m_1m_pr^3}\left[1+\frac{m_1}{m_2}-
\frac{Zm_1m_p}{2m_2^2\mu_h}\right]({\bf L}{\mathstrut\bm\sigma}_2)
\int_1^\infty\rho(s)ds e^{-2m_esr}\left(1+2m_esr\right)-
\end{equation}
\begin{displaymath}
-\frac{\alpha\mu_h(1+a_\mu)}{4m_1m_pr^3}\int_1^\infty\rho(s)dse^{-2m_esr}
\Bigl[4m_e^2s^2r^2\left({\mathstrut\bm\sigma}_1{\mathstrut\bm\sigma}_2-
({\mathstrut\bm\sigma}_1{\bf n})({\mathstrut\bm\sigma}_2{\bf
n})\right)+
\end{displaymath}
\begin{displaymath}
+(1+2m_esr)\left({\mathstrut\bm\sigma}_1{\mathstrut\bm\sigma}_2-
3({\mathstrut\bm\sigma}_1{\bf n})({\mathstrut\bm\sigma}_2{\bf
n})\right) \Bigr].
\end{displaymath}
The subsequent transformations of the diagonal matrix elements of
the operator (29), connected with the averaging over angles, can be
performed with the use of the formula (23) for the second term in
the square brackets of Eq.(29). Similar averaging for the first term
in the square brackets is:
\begin{equation}
\langle\delta_{ij}-n_in_j\rangle
=\frac{1}{5}(2\delta_{ij}+\overline{L_iL_j}+\overline{L_jL_i}).
\end{equation}
Then the contributions of the vacuum polarization in the first and
second orders of perturbation theory can be written as follows:
\begin{equation}
\Delta
E_1^{hfs}(2P_{1/2})=E_F\frac{\alpha}{18\pi}\int_1^\infty\rho(s)ds
\int_0^\infty xdx e^{-x\left(1+\frac{2m_es}{W}\right)}\times
\end{equation}
\begin{displaymath}
\times\left[\left(1+\frac{m_1}{m_2}-\frac{Zm_1m_p}{2m_2^2\mu_h}\right)\left(1+
\frac{2m_es}{W}x\right)+(1+a_\mu)\left(\frac{2m_e^2s^2x^2}{W^2}+1+\frac{2m_esx}
{W}\right)\right]=-65.21~~\mu eV,
\end{displaymath}
\begin{equation}
\Delta
E_1^{hfs}(2P_{3/2})=E_F\frac{\alpha}{18\pi}\int_1^\infty\rho(s)ds
\int_0^\infty xdx e^{-x\left(1+\frac{2m_es}{W}\right)}\times
\end{equation}
\begin{displaymath}
\times\left[\left(1+\frac{m_1}{m_2}-\frac{Zm_1m_p}{2m_2^2}\right)\left(1+
\frac{2m_es}{W}x\right)-\frac{(1+a_\mu)}{5}\left(\frac{8m_e^2s^2x^2}{W^2}+1+\frac{2m_esx}
{W}\right)\right]=-11.15~~\mu eV,
\end{displaymath}
\begin{equation}
\Delta
E_2^{hfs}(2P_{1/2})=E_F\frac{\alpha}{324\pi}\int_1^\infty\rho(s)ds
\int_0^\infty dx e^{-x\left(1+\frac{2m_es}{W}\right)}\times
\end{equation}
\begin{displaymath}
\times\int_0^\infty\frac{dx'}{x'^2}e^{-x'}g(x,x')
\left[2+\frac{m_1}{m_2}-\frac{Zm_1m_p}{2m_2^2\mu_h}+a_\mu\right]=-56.79~~\mu
eV,
\end{displaymath}
\begin{equation}
\Delta
E_2^{hfs}(2P_{3/2})=E_F\frac{\alpha}{324\pi}\int_1^\infty\rho(s)ds
\int_0^\infty dx e^{-x\left(1+\frac{2m_es}{W}\right)}\times
\end{equation}
\begin{displaymath}
\times\int_0^\infty\frac{dx'}{x'^2}e^{-x'}g(x,x')
\left[\frac{4}{5}+\frac{m_1}{m_2}-\frac{Zm_1m_p}{2m_2^2\mu_h}-\frac{a_\mu}{5}\right]=
-23.42~~\mu eV.
\end{displaymath}

\begin{table}
\caption{Hyperfine structure of $P$-wave energy levels in muonic
helium ion $(\mu^3_2He)^+$.}
\bigskip
\begin{tabular}{|c|c|c|c|}   \hline
Contribution to hyperfine& Numerical value &Numerical value &Reference, \\
splitting &of the contribution to &of the contribution to& equation  \\
    & $\Delta E^{hfs}(2P_{1/2})$,~$\mu eV$ &$\Delta E^{hfs}(2P_{3/2})$,~$\mu eV$ &     \\   \hline
Contribution of order $\alpha^4$ & -58356.61  & -24088.50
&\cite{p1,b1}, (24), (25)\\  \hline Muon AMM contribution & -33.29 &
6.66  &\cite{b1}, (24), (25)     \\   \hline Relativistic correction
of order $\alpha^6$  & -26.62    & -1.59 & (28)
\\   \hline
Contribution of order $\alpha^5$  &-65.21    &-11.15      & \cite{b1}, (31),(32)       \\
in the first order PT $\langle\Delta V^{hfs}_{VP}\rangle$    & & &
\\   \hline
Contribution of order $\alpha^5$  &          &    &(33),(34)    \\
in the second order PT  &-56.79   &-23.42    &  \\
$\langle\Delta V^C_{VP}\cdot\tilde G\cdot \Delta V^{hfs}\rangle$& &
&     \\   \hline
VP Contribution in the second &      &    & (4),(29)  \\
order PT of order $\alpha^6$  & -0.12 &-0.07    &    \\
$\langle\Delta V^C_{VP}\cdot\tilde G\cdot\Delta V^{hfs}_{VP}\rangle$&    &   &   \\  \hline
VP Contribution of $1\gamma$ interaction & -0.10    &  -0.01   &  (35)  \\
of order $\alpha^6$ $\langle\Delta V^{hfs}_{VP-VP}\rangle$& & &
\\   \hline
VP Contribution of $1\gamma$ interaction & -0.37     &-0.08   &  (36) \\
of order $\alpha^6$ $\langle\Delta V^{hfs}_{2-loop,VP}\rangle$& & &
\\   \hline
VP Contribution in the second & -0.01     &-0.004  & (11),(18),(21)  \\
order PT of order $\alpha^6$    &    &  &  \\
$\langle\Delta V^C_{VP-VP}\cdot\tilde G\cdot\Delta V^{hfs}\rangle$    &   &  &  \\  \hline
VP Contribution in the second &  -0.45    & -0.19   & (11),(19),(21) \\
order PT of order $\alpha^6$    &  &   &    \\
$\langle\Delta V^C_{2-loop,VP}\cdot\tilde G\cdot\Delta V^{hfs}\rangle$& & &
\\  \hline
Nuclear structure correction & -0.33 &  0.66 & (38) \\  \hline
Summary contribution  &  -58539.90   & -24117.69 &    \\   \hline
\end{tabular}
\end{table}

Two-loop vacuum polarization corrections to the hyperfine part of
the potential for the $l\not =0$ can be obtained with the aid of
relations (11) and (14). The results are
\begin{equation}
\Delta
V^{hfs}_{VP-VP}(r)=\frac{Z\alpha\mu_h}{2m_1m_2r^3}\left(\frac{
\alpha}{\pi}\right)^2\int_1^\infty\rho(\xi)d\xi\int_1^\infty\rho(\eta)d\eta
\frac{1}{\xi^2-\eta^2}\times
\end{equation}
\begin{displaymath}
\times\Biggl\{\left[1+\frac{m_1}{m_2}-\frac{Zm_1m_p}{2m_2^2\mu_h}\right]
({\bf L}{\mathstrut\bm\sigma}_2)\left[\xi^2(1+2m_e\xi r)e^{-2m_e\xi
r}- \eta^2(1+2m_e\eta r)e^{-2m_e\eta r}\right]-
\end{displaymath}
\begin{displaymath}
-\frac{1+a_\mu}{2}\Bigl[
\left({\mathstrut\bm\sigma}_1{\mathstrut\bm\sigma}_2-
3({\mathstrut\bm\sigma}_1{\bf n})({\mathstrut\bm\sigma}_2{\bf n})\right)
\left(\xi^2(1+2m_e\xi r)e^{-2m_e\xi r}-\eta^2(1+2m_e\eta r)
e^{-2m_e\eta r}\right)+
\end{displaymath}
\begin{displaymath}
+4m_e^2r^2\left({\mathstrut\bm\sigma}_1
{\mathstrut\bm\sigma}_2-3({\mathstrut\bm\sigma}_1{\bf n})
({\mathstrut\bm\sigma}_2{\bf n})\right)\left(\xi^4 e^{-2m_e\xi r}-
\eta^4 e^{-2m_e\eta r}\right)\Bigr]\Biggr\},
\end{displaymath}
\begin{equation}
\Delta
V_{2-loop,VP}^{hfs}(r)=\frac{Z\alpha\mu_h}{2m_1m_2r^3}\frac{2}
{3}\left(\frac{\alpha}{\pi}\right)^2\int_0^1\frac{f(v)dv}{1-v^2}e^{-\frac{2m_er}
{\sqrt{1-v^2}}}\times
\end{equation}
\begin{displaymath}
\times\Biggl\{\left[1+\frac{m_1}{m_2}-\frac{Zm_1m_p}{2m_2^2\mu_h}\right]
\left(1+\frac{2m_er}{\sqrt{1-v^2}}\right)({\bf L}
{\mathstrut\bm\sigma}_2)-\frac{1+a_\mu}{2}\times
\end{displaymath}
\begin{displaymath}
\times\left[\frac{4m_e^2r^2}{1-v^2}
\left({\mathstrut\bm\sigma}_1{\mathstrut\bm\sigma}_2-({\mathstrut\bm\sigma}_1
{\bf n})({\mathstrut\bm\sigma}_2{\bf n})\right)+
\left(1+\frac{2m_er}{\sqrt{1-v^2}}\right)
\left({\mathstrut\bm\sigma}_1{\mathstrut\bm\sigma}_2-3({\mathstrut\bm\sigma}_1
{\bf n})({\mathstrut\bm\sigma}_2{\bf n})\right)\right]\Biggr\}.
\end{displaymath}
Omitting further details of the calculations that can be performed
with the aid of a procedure similar to that which was used in the
derivation (17) and (20), we represent in Table II the numerical
values of the contributions to the energy spectrum that are
determined by the potentials (35) and (36). Yet another part of
two-loop corrections to the hyperfine structure in the second order
PT is shown in Fig.3. We also included in Table II numerical results
of the contributions from these amplitudes to hyperfine structure of
P-wave states.

\begin{figure}[htbp]
\includegraphics{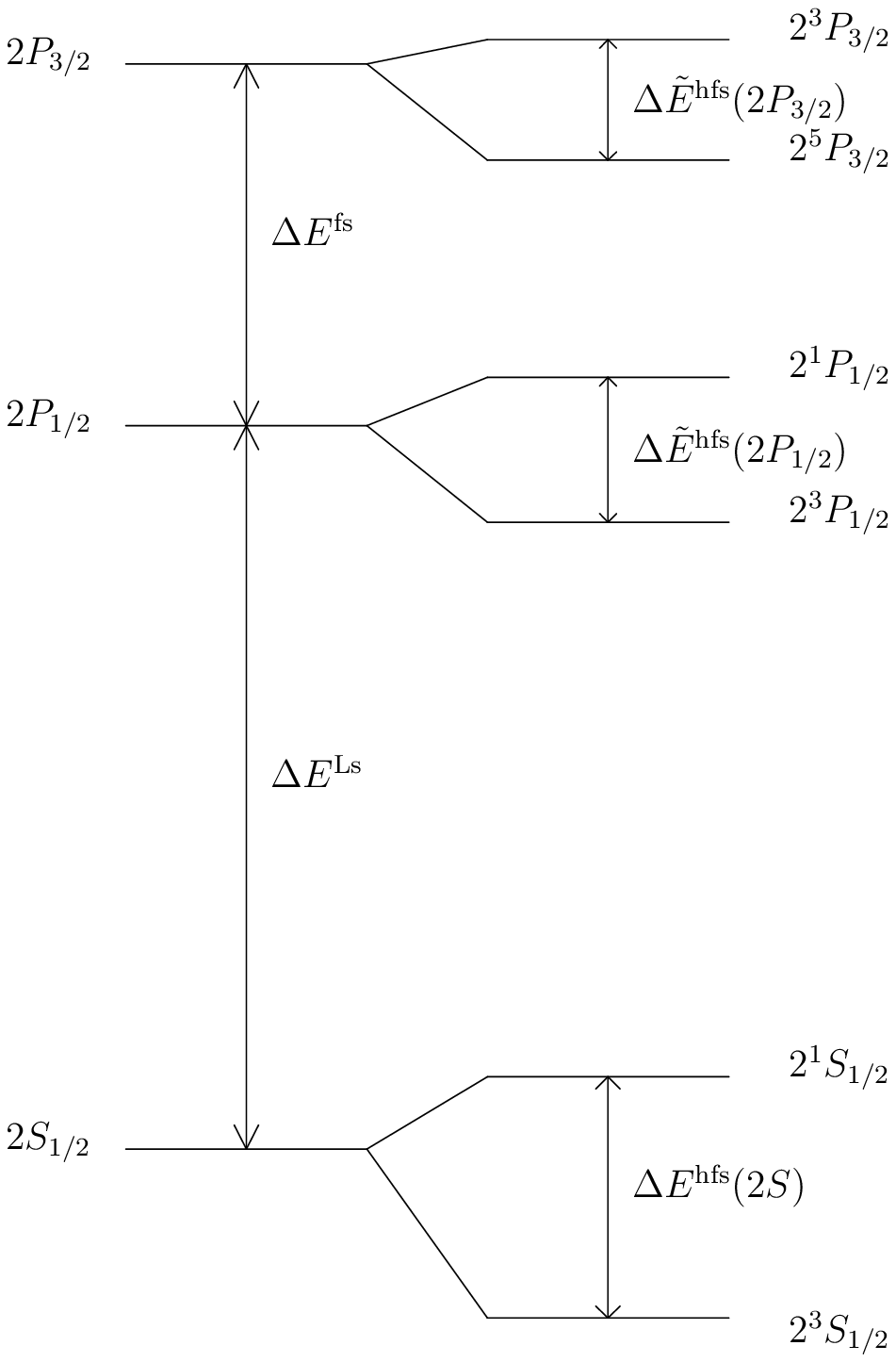}
\centering \caption{The structure of $S$-wave and $P$-wave energy
levels in muonic helium ion $(\mu ^3_2He)^+$ for the $n=2$.}
\end{figure}

Nuclear structure effects play significant role in the precise
calculation of the hyperfine structure in muonic atoms \cite{m1,m2}.
In order to calculate it in the case of P-wave levels we can expand
the helion magnetic form factor over relative momenta and obtain the
hyperfine part of the interaction operator in the momentum
representation, which is proportional to magnetic mean-square radius
$<r_M^2>$ of the helion:
\begin{equation}
\Delta V^{hfs}_{str}({\bf k})=-\frac{\pi
Z\alpha(1+a_\mu)<r_M^2>}{6m_1m_2}\left[({\mathstrut\bm\sigma}_1
{\mathstrut\bm\sigma}_2){\bf k}^2-({\mathstrut\bm\sigma}_1{\bf k})
({\mathstrut\bm\sigma}_2{\bf k})\right],
\end{equation}
Averaging this operator over the wave functions $\psi_{2P}$ we
obtain the following contribution to the hyperfine structure:
\begin{equation}
\Delta
E^{hfs}_{str}=\frac{\mu^5(Z\alpha)^6(1+a_\mu)<r_M^2>}{144m_1m_2}
\frac{[F(F+1)-j(j+1)-\frac{3}{4}][j(j+1)-\frac{5}{4}]}{j(j+1)}.
\end{equation}
Corresponding numerical values for the hyperfine splittings of
$2P_{1/2}$ and $2P_{3/2}$ states are included in Table II. We assume
that the values of the charge and magnetic mean-square radii
coincide and take the value of $^3_2He$ charge radius
$r_E=1.9642(11)$ fm \cite{3he}.

The off-diagonal matrix element has an important role to attain the
high accuracy of the calculation of the $P$-wave levels in muonic
helium ion. We present its general structure as follows:
\begin{equation}
\gamma=\langle 2^3P_{1/2}|\Delta
V^{hfs}|2^3P_{3/2}\rangle=E_F\left(-\frac{\sqrt{2}}{48}
\right)\left[1-a_\mu+\frac{2m_1}{m_2}-\frac{Zm_1m_p}{m_2^2\mu_h}+
\frac{m_1^3}{\mu^3}C_{rel}(Z\alpha)^2+C_{VP}\alpha\right],
\end{equation}
where for the sake of simplicity, we have restricted ourselves to
terms of fifth order in $\alpha$ in considering vacuum polarization
effects, terms of fifth and higher orders in the muon anomalous
magnetic moment, relativistic effects of order $(Z\alpha)^6$ and
recoil effects. The first three terms in the right-hand side of Eq.
(37) result from employing the potential (21). In the Dirac's
theory, relativistic corrections are determined by off-diagonal
radial integrals:
\begin{equation}
R_{\frac{1}{2}\frac{3}{2}}=\int_0^\infty\left(g_{\frac{1}{2}}(r)f_{\frac{3}{2}}(r)
+g_{\frac{3}{2}}(r)f_{\frac{1}{2}}(r)\right)dr.
\end{equation}
By using the explicit expressions for the wave functions
$f_{1/2,3/2}(r)$ and $g_{1/2,3/2}(r)$ \cite{BS} in order to
calculate this integral, we obtain the coefficient $C_{rel}=9/16$.
In order to obtain the vacuum polarization correction in $\gamma$ we
use the potential (29). Then we have to calculate the matrix
elements of the following operators:
\begin{equation}
T_1=({\bf L}{\mathstrut\bm\sigma}_2),~T_2=
\left[{\mathstrut\bm\sigma}_1{\mathstrut\bm\sigma}_2-3({\mathstrut\bm\sigma}_1
{\bf n})({\mathstrut\bm\sigma}_2{\bf n})\right],~T_3=
\left[{\mathstrut\bm\sigma}_1{\mathstrut\bm\sigma}_2-({\mathstrut\bm\sigma}_1
{\bf n})({\mathstrut\bm\sigma}_2{\bf n})\right].
\end{equation}
Upon averaging over angles with the aid of expressions (23) and
(30), they can be written in terms of the $6j$ - symbols as
\begin{equation}
\langle T_3\rangle=-\langle T_2\rangle=-\langle T_1\rangle= =6\hat
j\hat j' \Biggl\{{{l~F~1}\atop{\frac{1}{2}~\frac{1}{2}~j}}
\Biggr\}\Biggl\{{{l~F~1}
\atop{\frac{1}{2}~\frac{1}{2}~j'}}\Biggr\}=\frac{2\sqrt{2}}{3},
\end{equation}
where the value of the total momentum $F=1$ $({\bf F}={\bf s}_2+{\bf
J})$, $l=1$, $\hat j=\sqrt{2j+1}$, $\hat j'=\sqrt{2j'+1}$, and the
numerical values of the $6j$ - symbols are borrowed from
\cite{Sobelman}. As a result the vacuum polarization contributions
to the off-diagonal matrix element (39) in the first and second
orders of perturbation theory have the form:
\begin{equation}
\gamma_1=\langle 2^3P_{1/2}|\Delta V^{hfs}_{VP}|2^3P_{3/2}\rangle=
E_F\left(-\frac{\sqrt{2}}
{72}\right)\frac{\alpha}{\pi}\int_1^\infty\rho(s)ds\int_0^\infty
xdxe^{-x\left( 1+\frac{2m_es}{W}\right)}\times
\end{equation}
\begin{displaymath}
\times\left[\left(1+\frac{m_1}{m_2}-\frac{Zm_1m_p}{2m_2^2\mu_h}\right)\left(1+
\frac{2m_es}{W}x\right)-\frac{1+a_\mu}{2}\left(1+\frac{2m_es}{W}x-
\frac{4m_e^2s^2}{W^2}x^2\right)\right]=9.49~\mu eV,
\end{displaymath}
\begin{equation}
\gamma_2=\langle 2^3P_{1/2}|\Delta V_{VP}^C\cdot \tilde G\cdot
\Delta V^{hfs}_{B}|2^3P_{3/2}\rangle=
E_F\left(-\frac{\sqrt{2}}{2592}\right)\frac{\alpha}{\pi}
\left[1+\frac{2m_1}{m_2}-\frac{Zm_1m_p}{m_2^2\mu_h}-a_\mu\right]\times
\end{equation}
\begin{displaymath}
\times\int_1^\infty\rho(s)ds \int_0^\infty
xdxe^{-x\left(1+\frac{2m_es}{W}\right)}
\int_0^\infty\frac{dx'}{x'^2}e^{-x'}g(x,x')=5.33~\mu eV.
\end{displaymath}
The total numerical value of the matrix element (39) is
$\gamma=5497.28~\mu eV$. It leads to the shift of the hyperfine
splittings of the energy levels $2^3P_{3/2}$ and $2^3P_{1/2}$ by the
value $\delta=173.0~\mu eV$.

\section{Summary and conclusion}

In the present study we have calculated QED effects in the fine and
hyperfine structure of the $2P_{1/2}$ and $2P_{3/2}$ energy levels
in muonic helium ion $(\mu^3_2He)^+$. We have considered the
electron vacuum polarization contributions of orders $\alpha^5$,
$\alpha^6$, recoil corrections, relativistic effects of order
$\alpha^6$ and nuclear structure corrections. The numerical values
of the contributions are presented in Tables I and II. In these
Tables we give the references to other papers also devoted to the
investigation of the structure of $P$-wave levels in the hydrogenic
atoms.

Let us summarize the basic particularities of the calculation performed
above.

1. Special attention in our investigation has been concentrated on
the vacuum polarization effects. For this purpose we obtain the
terms of the interaction operator in muonic helium ion which contain
the one-loop and two-loop vacuum polarization corrections.

2. In each order in $\alpha$ we have taken into account recoil
effects in the terms proportional to $m_1/m_2$. The experimental
values of the muon and helion magnetic moments are used \cite{MT}.

3. The calculation of the relativistic corrections to the diagonal
and nondiagonal matrix elements both for the fine and hyperfine
structure intervals is performed on the basis of the Dirac equation.

Total numerical values for the fine structure interval $\Delta
E^{fs}$ (3) and hyperfine splitting intervals of $2P_{1/2}$ and
$2P_{3/2}$ states are presented in Tables I,II. Taking also into
account our calculation of the mixing of the $2^3P_{3/2}-$ and
$2^3P_{1/2}-$ wave energy levels (the correction $\delta$), we find
that these values of hyperfine splittings change by
$\delta=173.0~\mu eV$: $\Delta \tilde E^{hfs}(2P_{1/2})=\Delta
E^{hfs}(2P_{1/2})-\delta=-58712.90~\mu eV$, $\Delta \tilde
E^{hfs}(2P_{3/2})=\Delta E^{hfs}(2P_{3/2})-\delta= -24290.69~\mu
eV$. The theoretical error of the obtained results is determined by
the contributions of higher order and amounts up to $10^{-6}$.
Previously, the $(2S-2P)$ transition energies for muonic helium ion
were studied in \cite{b1,b2,b3}. Considering the obtained in
\cite{b2,b3} numerical results for different transitions $(2s-2p)$
we find that the fine splitting and the hyperfine splitting
intervals for the states $2P_{1/2}$ and $2P_{3/2}$ in this paper are
equal: $\Delta E^{fs}=145.0~meV$, $\Delta
E^{hfs}(2P_{1/2})=58.3~meV$, $\Delta E^{hfs}(2P_{3/2})=24.5~meV$.
So, the results of our work agree and refine the previous
calculations performed in \cite{b2,b3} via taking into account
higher order effects. They can be considered as a reliable estimate
for the fine and hyperfine structure intervals for the $P$- levels
in muonic helium ion $(\mu ^3_2He)^+$. The disposition of the
$P$-wave energy levels is shown in Fig.4. Taking into account the
value of the Lamb shift $(2S-2P)$ from \cite{b1}, our result for the
hyperfine splitting of $2S$ state from \cite{m1} and the numerical
results obtained in this work, we find new $(2S-2P)$ transition
energies, which are presented in Table III.

\begin{table}
\caption{$(2S-2P)$ transition energies for muonic helium ion
$(\mu^3_2He)^+$.}
\bigskip
\begin{tabular}{|c|c|c|}   \hline
Transition &  Energy (meV)  &   \cite{b2}    \\   \hline
$^1S_{1/2}-^3P_{1/2}$   &   1167.33  &  1167.3   \\   \hline
$^1S_{1/2}-^3P_{3/2}$   &   1342.19  &  1342.3   \\   \hline
$^3S_{1/2}-^1P_{1/2}$   &   1392.66  &  1392.4   \\   \hline
$^3S_{1/2}-^3P_{1/2}$   &   1333.95  &  1334.1   \\   \hline
$^3S_{1/2}-^3P_{3/2}$   &   1508.80  &  1509.1   \\   \hline
$^3S_{1/2}-^5P_{3/2}$   &   1484.51  &  1484.6   \\   \hline
\end{tabular}
\end{table}

\begin{acknowledgments}
The work is performed under the financial support of the Federal
Program "Scientific and pedagogical personnel of innovative
Russia"(grant No. NK-20P/1)
\end{acknowledgments}

\end{document}